\documentclass{aastex}
\usepackage{emulateapj5}
\usepackage{subfigure}

\shorttitle{Crab-like SNR G21.5$-$0.9  at millimeter wavelengths}
\shortauthors{Bock, Wright, \& Dickel}

\begin{document}
\slugcomment{ApJL, in press (accepted October 2, 2001)}

\title{The Crab-like Supernova Remnant G21.5$-$0.9 at Millimeter Wavelengths}

\author {Douglas C.-J. Bock, Melvyn C. H. Wright}
\affil{Radio Astronomy Laboratory, University of California, 
  Berkeley, CA 94720}
\email{dbock@astro.berkeley.edu, mwright@astro.berkeley.edu}
\and
\author{John R. Dickel}
\affil{Astronomy Department, University of Illinois at
  Urbana-Champaign, Urbana, IL 61801}
\email{johnd@astro.uiuc.edu}

  \begin{abstract}

    We present a BIMA image of the supernova remnant G21.5$-$0.9 at
    94~GHz with angular resolution $8.6''\times 4.5''$. On scales
    larger than our synthesized beam, our results do not indicate any
    radial or local variations in the acceleration or synchrotron loss
    processes for the relativistic particles emitting at radio
    wavelengths. However, the FWHM size of the radio remnant is
    significantly greater than that of the X-ray remnant. Either any
    low-frequency spectral break is distributed across the remnant, or
    more probably the break frequency is in fact higher than $\sim
    100$ GHz.

\end{abstract}
 
\keywords{ISM: individual (G21.5$-$0.9) --- radio continuum:
  ISM --- stars: neutron --- stars: winds --- supernova remnants}
   
\section{Introduction}

Pulsar-powered nebulae have a characteristic radio spectrum; powering
by the embedded neutron star, whether observed or not, creates a
relatively flat synchrotron spectrum with a spectral index $\alpha$
($S_\nu \propto \nu^\alpha$) between 0 and $-0.4$ at low frequencies.
Above this regime one or more breaks occur in the spectrum as
synchrotron losses, reduced energy from the pulsar with time, and
expansion of the nebula play a role.  At high frequencies, synchrotron
losses will dominate the spectrum and a break will occur such that the
higher frequency spectral index should be steeper by 0.5 \citep{kar62,
  rey84}.  When the expansion, pulsar decay, and synchrotron losses
are all important, an intermediate region with an intermediate slope
will be present in the spectrum.  The break points will generally
diverge with time \citep{wol97}.  In addition, non-thermal X-rays may
come from the same or different parent populations of relativistic
electrons.

To determine the parameters of the Crab-like SNRs, it is necessary to
measure the frequencies at which the spectral breaks occur and the
actual spectral indices in the different spectral components. Also,
because the expansion rate and the magnetic fields may vary with
position within a remnant, it is important to image in the
region of the break frequency to look for possible variations across the
source.  The lower break frequency generally falls somewhere between
$10^9$ and $10^{13}$ Hz \citep{sal89b,gre92}. The Crab itself has a
break frequency of about $10^{13}$ Hz, well above the mm band
\citep{str92}.  However, measuring spatial variations in the mm-wave
region of the spectrum is a difficult task which has not been done for
any Crab-like SNR, since the synchrotron emission is generally weak
and, while the sources are often 
extended, high resolution observations are needed to resolve structure 
across them. 

G21.5$-$0.9 is one candidate
which can be studied in detail.  Integrated flux density measurements
indicate a break frequency near 50~GHz and a total flux density of
4~Jy at 84.2~GHz \citep{sal89b}.  The object has been imaged with $7''
\times 4''$ resolution at both 4.9~GHz \citep{bec81} and 22.3~GHz
\citep{fue88}.  As expected, these two low-frequency images appear the
same with a filled center and some structure.  In addition, the center
has very low polarization but is surrounded by a ring of highly
polarized emission with a radial magnetic field.  This may well
indicate different conditions in the interior and around the outside
of this SNR.

Recent X-ray observations have painted a very different picture at
higher energies \citep{sla00,war01}. Here the emission is seen over a
greater area, and has several distinct components: an unpulsed,
compact X-ray source, an extended core corresponding to the radio
nebula, and a faint outer shell of radius about 150 arcsec which is
not observed in the radio.

In this letter we present the first interferometric image above the
previously suggested break frequency using the BIMA\footnote{The BIMA
  array is operated by the Berkeley-Illinois-Maryland Association
  under funding from the National Science Foundation} array at 94~GHz.
We compare this new image with images at 5 GHz, 22 GHz and
0.1--10~keV to look for structures corresponding to the spectral break
seen in the integrated flux density.

\section{Observations}
\label{sec:obs}

Observations were obtained with the B, C, and D configurations of the
BIMA array \citep{wel96} between 1998 October and 1999 March.  To
achieve suitable $uv$ coverage, we used the technique of
multi-frequency synthesis over both sidebands of the local
oscillator. We measured two orthogonal polarizations to ensure
accurate total intensities.  In order to image fully the extended
structure, and to assist with the recovery of low spatial frequency
information, we imaged the source with a 7-point hexagonal mosaic.
The primary beam of the 6.1-m antennas is Gaussian with FWHM $2.13'$
and $1.74'$ respectively at 90, and 110 GHz.

The quasar 1743$-$038 was observed as a phase calibrator at 25 to 30
minute intervals, after 3 complete cycles of 7 pointings.  The antenna
gains were determined from observations of planets at short antenna
separations; the uncertainty in the flux density scale is 10\%. All
data were obtained in an 800 MHz bandwidth using a digital correlator
with 256 frequency channels.  Single sideband system temperatures
during the observations were between 170 and 450 K (scaled to outside
the atmosphere).

\begin{figure}[p]
\centering
\subfigure[]{\includegraphics[width=8cm, angle=0]{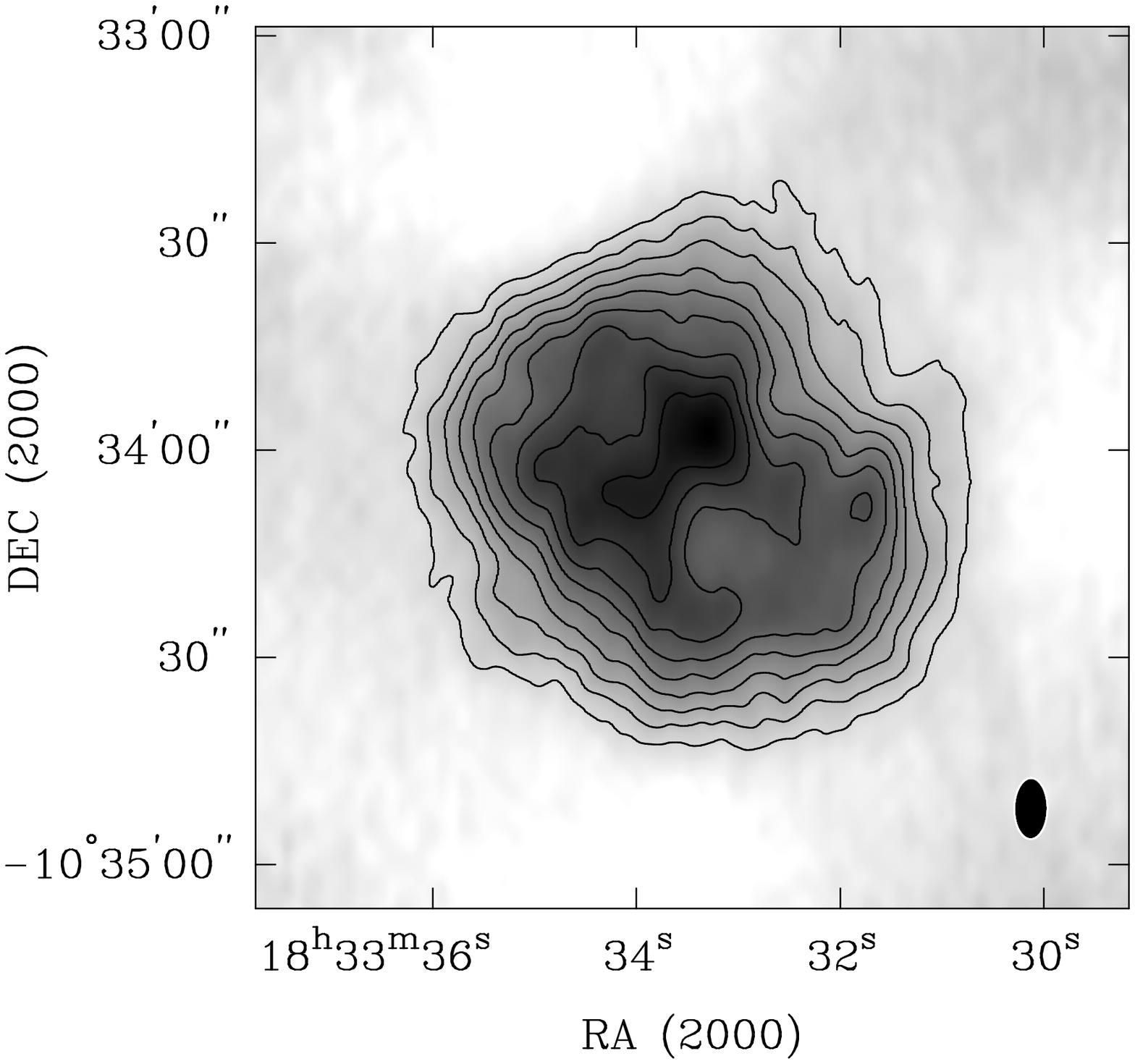}}
\subfigure[]{\includegraphics[width=8cm, angle=0]{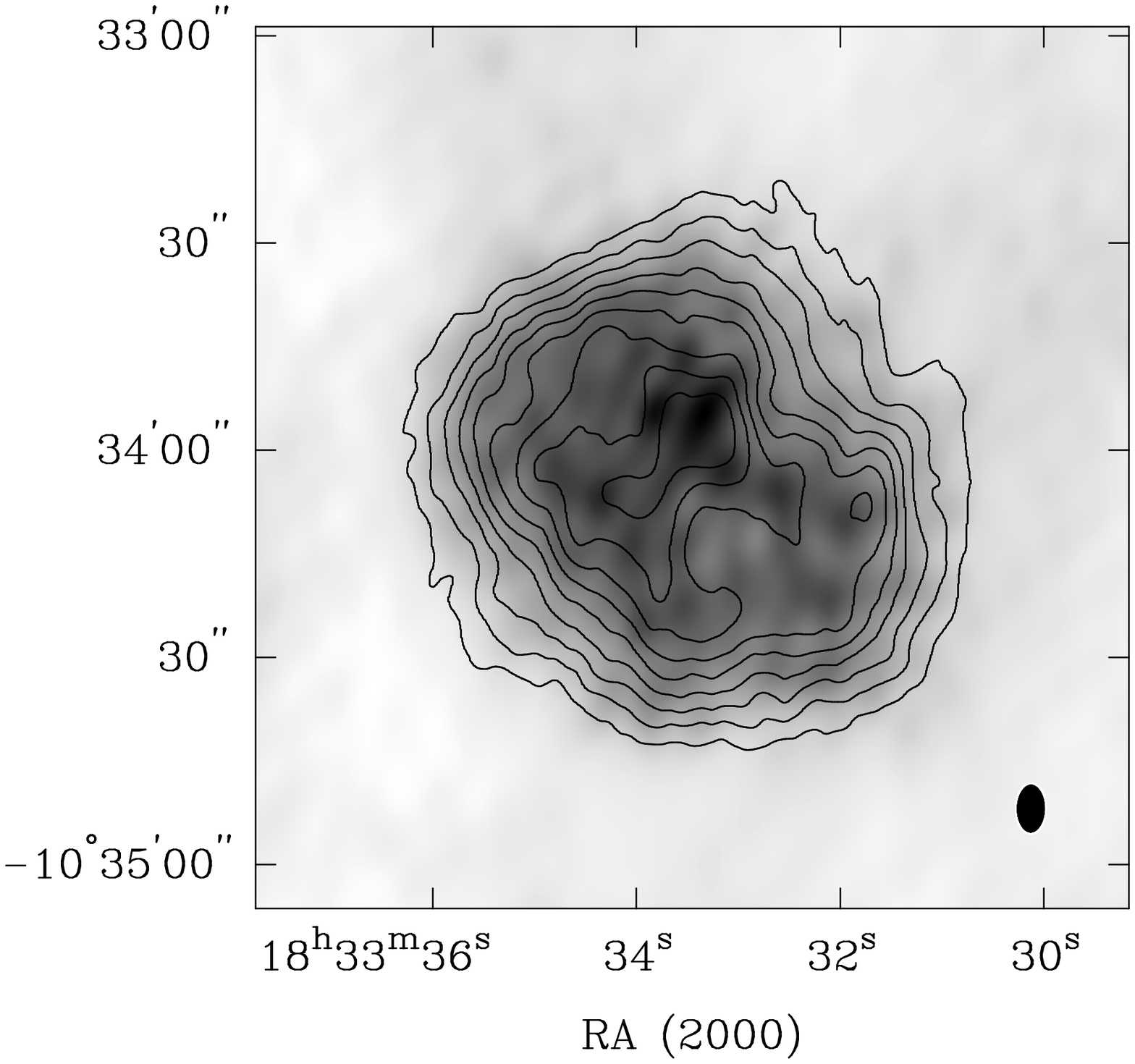}}

\figcaption[bock_fig1a.eps,bock_fig1b.eps]{94 GHz mosaic image of
  G21.5$-$0.9 (contours) overlaid on (a) greyscale of 94 GHz data and
  (b) greyscale of 22 GHz Nobeyama data. The contour interval is 7 mJy
  beam$^{-1}$, with the lower contour at 7 mJy beam$^{-1}$. The 22~GHz
  image has been normalized to the BIMA intensity scale (see text).
  The restoring beams for each image are
  shown.\label{fig:grey}}
\end{figure}

Imaging was done with the MIRIAD software package (\citealt*{sau95})
using a maximum entropy mosaicing algorithm (\citealt*{sau96}), in
which the multiple pointing centers were combined into a mosaiced
image corrected for the primary beam. To assess imaging reliability
for a complicated source we must consider image fidelity (i.e.\ how
well the image represents the real source brightness distribution),
rather than the more familiar measures of thermal noise level or image
dynamic range. To do this we re-calibrated and re-imaged G21.5$-$0.9
many times by choosing
subsets of the data, and using
a variety of $uv$ weighting schemes (natural, uniform and robust).
The RMS difference between these images was consistent with the
thermal noise at 94 GHz (3 mJy beam$^{-1}$).  The peak differences
between these images was 12 mJy beam$^{-1}$, so differences between
images corresponding to as much as 1.5 contours on
figure~\ref{fig:grey} may not be significant.  We use these empirical
estimates of the RMS in evaluating the results.

\section{Results}

Figure~\ref{fig:grey}\notetoeditor{Figure 1(a) and (b) should be
  reproduced side by side if used in the current format; if not, the
  axis labels will need to be redone. Please let me know.} shows the
maximum entropy image with mean frequency 93.7 GHz.  The convolving
half-power beamwidth was $8.6'' \times 4.5''$.  This image includes
measurements on spatial scales up to 2 arcmin.  The 94 GHz image is
similar to the 5 GHz \citep{bec81} and 22 GHz \citep{fue88} images
with comparable angular resolution, but does not reproduce the
detailed small-scale structure seen at 22 GHz, even after convolution
to the resolution ($8''$) used for detailed studies by \citet{fue88}.
If real, these features should be above the noise level in our image.
It may be that the `CLEAN' algorithm used by them produced a mottled
appearance in the smooth emission \citep{cor99}.  The three
interferometric images also differ somewhat from the single dish 90
GHz image \citep{sal89a}: the secondary peak to the north east in that
image is not seen with the interferometers. Given the difficulties
inherent in single-dish imaging, we do not consider this discrepancy
significant.

The total flux density recovered in the final image is 3.5 Jy measured
within a radius of $45''$. The total flux densities measured on the
individual images made from the 90 and 110 GHz data separately (3.4 Jy
and 3.6 Jy) are consistent within the calibration errors of about
10\%.  Omitting the most compact array configuration reduces the
measured total flux density, to 3.0 Jy, as expected. We conclude that
the final image recovers all the emission corresponding to the known
radio extent, with a total flux density $3.5\pm0.4$ Jy at 94 GHz, in agreement
with the single-dish result at 91 GHz of $3.8\pm0.4$ Jy
\citep{sal89a}. However, we note that the total flux density has been
estimated in the deconvolution and has not been directly measured by
the interferometer.

\subsection{Radial profile analysis}

\begin{figure}[p]
\centering
\includegraphics[width=8cm]{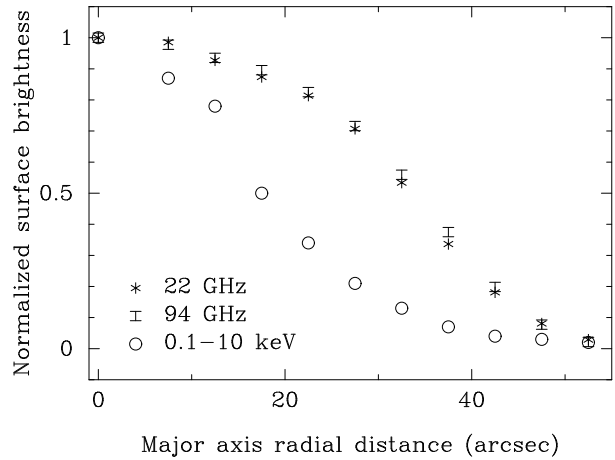}

\figcaption[bock_fig2.eps]{Radial profiles of G21.5$-$0.9 at 22~GHz
  (Nobeyama), 94~GHz (BIMA), and in the 0.1--10 keV band
  (\emph{Chandra} HRC).  The 94-GHz symbol sizes represent the
  uncertainty due to imaging errors.\label{fig:mw_prof}}
\end{figure}

In order to evaluate the spectral index distribution accurately, we
must match the data as closely as possible, not only in angular
resolution but also in the spatial sampling of the images. One
possibility is to compare completely sampled images at each frequency.
We have compared the normalized surface brightness in elliptical
annuli of the 94~GHz, 22~GHz \citep{fue88} and 0.1--10~keV
images.\footnote{The X-ray data used were from HRC image \#1406 (1999
  October 25) in the \emph{Chandra} archive.} The 5~GHz maps
\citep{bec81} were not available in digital format. The best-fit
center for the analysis was determined empirically to be
$18\mathrm{^h}33\mathrm{^m}33\fs4$, $-10\degr34\arcmin04\arcsec$
(J2000). The ellipses had position angle 45\degr\ and axial ratio 0.6.
To make the comparison we subtracted base levels of 5000~K, $-$2~mJy
beam$^{-1}$ and 2 counts respectively from the 22~GHz, 94~GHz and
X-ray images, and then normalized the data at the center, after
removing the contribution of the compact X-ray source. The comparison,
using annuli of width 5 arcsec, is plotted in
figure~\ref{fig:mw_prof}.

The errors in our radial profile at 94 GHz were estimated empirically
from the RMS variation between the various images discussed in
\S\ref{sec:obs}. Using 1.5 arcsecond bins, the RMS variation between
the derived radial profiles was generally less than 1.5\% of full
scale. The error bars on figure~\ref{fig:mw_prof} indicate this
uncertainty.

The two radio datasets agree within our estimate of the error at all
but two points on the plot. Taking into account the sensitivities of
the images and the `peak' errors in the imaging process
(\S\ref{sec:obs}), we do not find these discrepancies to be
significant. Thus we can find no significant radial differences
between the radio images at 22 and 94~GHz. A similar analysis was made
from the contour maps of the 5 and 94~GHz data using graphical
techniques. These were also found not to differ significantly.

By contrast, the X-ray shape of the plerion remnant differs
significantly from the radio shape.  At a radius of 30\arcsec, the
radio surface brightness is two-thirds of the central value, whereas
the X-ray surface brightness is only 15\% of its peak. The major axis
FWHM sizes of the radio and X-ray remnants are 68\arcsec\ and
35\arcsec\ respectively.

\subsection{Search for a radio shell}

G21.5$-$0.9 has recently been shown to have faint X-ray emission
extending well beyond the radio \citep{sla00,war01}. This may be the
signature of the SNR shell; in this case a radio counterpart would be
expected.  \citeauthor{sla00} report that they have been unable to
detect this shell at 1~GHz to a surface brightness limit of $4\times
10^{-21}$~W~m$^{-2}$~Hz$^{-1}$~sr$^{-1}$ (30~mJy~arcmin$^{-1}$). Any
emission with a canonical shell-type spectral index between $-0.5$ and
$-1$ would be far below our detection limit. Emission on the scales
\citeauthor{sla00} expected from the X-ray brightness distribution
would be heavily resolved except on the shortest of our baselines.
Modeling the X-ray emission with $60''$--$130''$ diameter Gaussian,
disk and shell distributions, we generated $uv$ data for the most
compact array configuration used.  The expected visibility of the
extended emission ranged from 5 to 50\% for these models.  After
tapering our data to a $1'$ beam, we were unable to detect any
emission to a ($1\sigma$) level of 10 mJy beam$^{-1}$ corresponding to
a source brightness limit of 20 to 200 mJy~arcmin$^{-1}$ for the above
models.

\section{Discussion}

The virtually identical radial brightness distributions above and
below the apparent break frequency given by \citet{sal89a} indicate no
change in the spectral index with radius.  This would appear to rule
out any change in magnetic field strength, and thus break frequency
with radius and also any variation in diffusion of particles
throughout the remnant. Alternatively the break may actually be above
the frequency of our observations. We discuss both possibilities
below, but first we revisit the questions of the age and distance of
G21.5$-$0.9.

The most reliable distance estimates for G21.5$-$0.9 have been minimum
distances obtained kinematically from hydrogen absorption
measurements. These estimates have ranged from 4.4 to 5.5~kpc. The
best determination of $v_{\rm LSR}$ remains that of \citet{dav86}
($v_{\rm LSR} = 65$ km s$^{-1}$).  We use $R_0 = 8.0$~kpc
\citep{rei93} and the Galactic rotation model of \citet{bra93} to
obtain a minimum distance to G21.5$-$0.9 of 4.4~kpc.  At that distance
the major axis of about $100''$ is 2.1 pc. See \citet{dav86} for a
discussion of the maximum distances sustainable from a kinematic
analysis.

There is no reliable age estimate for G21.5$-$0.9. While it is
physically smaller than the other Crab-like SNRs studied, it is much
fainter than the Crab and also has a low frequency for the break in
the spectrum, implying that the pulsar either did not provide as much
energy, or has had time to decay.  G21.5$-$0.9 may be the result of a
supernova in a very dense part of the Galaxy, or its progenitor may
have had a relatively weak stellar wind. The radio size alone
indicates that the remnant could be young compared to the historical
Crab-like SNRs (the Crab Nebula and 3C58). But the larger size of
the faint outer X-ray component indicates a greater age. A
reasonable estimate for the age could be 1000 years, although this
may be substantially in error.

\subsection{Spectral break below 100 GHz}

If the break frequency is below our observing frequency and particles
are injected locally by the neutron star as the nebula expands, we
would expect to see some spectral steepening at the edge of the
remnant, where electrons with shorter lifetimes would be
under-abundant. This would show up as a smaller size, as is seen in
X-rays but not in the radio.  Applying the standard minimum energy
requirement (equipartition), the average magnetic field for
G21.5$-$0.9 at our adopted minimum distance is 460$\mu$G
\citep{sla00}. This is comparable to estimates of the field in the
Crab Nebula obtained by this and other methods \citep{hes96}. The
conventional spectral break due to synchrotron losses in a field $B$
mG occurs at age $\sim 40000B^{-1.5}\nu^{-0.5}$ yr at frequency $\nu$
GHz. For $\nu_{\rm b} < 100$ GHz the estimated age of the break would
be $\gtrsim 13,000$ yr, larger than seems likely for this
small-diameter object,
unless it was formed in a very dense medium. Magnetic fields of
several mG are required to make the synchrotron lifetime comparable to
the age of most known Crab-like remnants exhibiting only a naked
plerion. This supports the view that this spectral break is above our
observing frequency.

Another explanation of the observed spectrum, discussed by
\citet{gre92}, is that in 3C58 and G21.5$-$0.9 the pulsar's excitation
has stopped and so the decay of the older electrons has proceeded
further and caused the steeper high frequency spectrum. If the
relativistic electrons do diffuse throughout the remnant, this could
allow the same structure above and below the break frequency. 
A related explanation for the low frequency of the spectral break and
the steeper spectrum above the break comes from a suggestion that
Crab-like SNRs might evolve to be
more like G21.5$-$0.9 when the pulsar input 
decays; young electrons observed in X-rays experience a weaker 
magnetic field 
when they are injected into the pulsar wind nebula 
than did the older ones still seen at radio wavelengths 
\citep{wol97}.
We note that 3C58
and G21.5$-$0.9 both have lower break frequencies and steeper spectral
indices above the break than remnants with active pulsars such as the
Crab Nebula and $0540-693$ (\citealt{gre2000}\footnote{Available at
http://www.mrao.cam.ac.uk/surveys/snrs.}, \citealt{man93}).

In summary, if the break is below our observing frequency, and due to
synchrotron losses, we would expect the spectral index to steepen with
distance from the remnant's center. Since this is not observed, either
the electrons were not injected with a simple power law and perhaps
have an intrinsic spectral flattening towards longer wavelengths, or
the radio-emitting particles have diffused throughout the remnant.

\subsection{Spectral break above 100 GHz}

The above discussion and the X-ray results seem to favor the
conclusion that the break is indeed above approximately 100 GHz.
Perhaps the slightly lower integrated flux densities in the 50--100
GHz range could be accounted for by different background levels chosen
by different observers, particularly if there is any faint radio
counterpart corresponding to the extended X-ray emission.

\citet{gal98} have measured the spectrum of G21.5$-$0.9 at infrared
wavelengths and find evidence for an X-ray to radio change in spectral
index of $-1.0$, which could indicate a break frequency much higher
than previously thought.  Extrapolation of the X-ray spectra
\citep{sla00,war01} to meet the extrapolation of a single power law
fit to all the radio data or just to the points below 30 GHz tends to
support this conclusion.  We note that a recent report of an
integrated flux density of 3.9 Jy for the entire SNR at a frequency of
230 GHz by \citet{ban01} also supports a single power law for the
radio spectrum at least beyond that frequency.  Sub-millimeter and
further infra-red observations are required to settle the question.

Overall, the observational data better support a higher break
frequency due to simple synchrotron losses, with spatial effects seen
in X-rays but not at radio wavelengths.  In this case, the particles
on the outer edges of the remnant should have been
injected/accelerated first, and are the ones showing evidence that the
highest energy particles have begun to decay.

\subsection{The X-ray morphology}

The radio data do not address the question of why there are distinct
breaks in the X-ray morphology between the central peak and the core
and again between the core and the plateau. The former characteristic
is similar to that seen in 3C58 \citep{hel95} but the latter appears
to be a unique property of G21.5$-$0.9.  One possibility is that there
could have been multiple events or injection epochs.  Although the
entire remnant seems to be pulsar-powered, expansion into different
ISM may affect the way the material responds to the pulsar
stimulation.  Alternatively, the pulsar may have a nebula or a
companion which could change its effect on its surroundings.  The
breaks in the X-ray spectrum may also be due to different decay rates
for the emitting particles in the core and the plateau. The recent
X-ray data argue against any thermal component \citep{saf01}.

\section{Conclusion}

Using BIMA observations at 94~GHz we have found G21.5$-$0.9
morphologically to have very similar structure at millimeter and
longer radio wavelengths, indicating no change in spectral index with
radius (although we have not been able to reproduce all the
small-scale structure seen at 22~GHz).  Particles emitting at radio
wavelengths either do not show evidence of a variation in lifetime
with distance from the assumed pulsar, or are well-diffused throughout
the remnant.  Since radial variations in the spectrum are seen at
X-ray wavelengths, the most likely scenario is that the break
frequency is above the frequency of the present observations.
Measurements at higher frequencies will most likely support this
picture. If they instead confirm a break frequency well below 94~GHz,
the break is not likely to be due to the limited lifetimes of the
higher energy electrons, but to some other mechanism.

\acknowledgements

This work was supported in part by NSF Grant AST-9981308 to the
University of California, and AST-9981363 to the University of
Illinois. We thank W. Reich for providing the 22~GHz image in
electronic form, and E.\ Amato, B.\ Gaensler, and M.\ Salvati for
useful discussions.

\begin{thebibliography}{}

\bibitem[Bandiera et al.(2001)]{ban01} Bandiera, R., Neri, R.\ \& Cesaroni, 
R., 2001, in AIP Conf.\ Ser.\ 565, Young Supernova Remnants, ed.\ S.\ Holt 
\& U.\ Huang (New York: AIP), 329 
 
%  4.885~GHz image.
\bibitem[Becker \& Szymkowiak(1981)]{bec81} Becker, R., \& Szymkowiak,
  A.\ 1981, ApJ, 248, l23

% rotation curve for the Galaxy
\bibitem[Brand \& Blitz(1993)]{bra93} Brand, J.\ \& Blitz, 
L.\ 1993, \aap, 275, 67 

\bibitem[Cornwell, Braun, \& Briggs(1999)]{cor99} Cornwell, 
T., Braun, R., \& Briggs, D.\ S.\ 1999, in ASP Conf.\ Ser. 180, Synthesis 
Imaging in Radio Astronomy II, ed.\ G.\ B.\ Taylor, C.\ L.\ Carilli, \&
R.\ A.\ Perley (San Francisco: ASP), 151 

\bibitem[Davelaar et al.(1986)Davelaar, Smith \& Becker]{dav86} Davelaar, 
J., Smith, A.\ \& Becker, R.\ H.\ 1986, \apj, 300, L59 

% 22 GHz image
\bibitem[F\"urst et al.(1988)]{fue88}
F\"urst, E., Handa, T., Morita, K., Reich, P., Reich, W., \& Sofue, Y.\ 1988,
PASJ, 40, 347

\bibitem[Gallant \& Tuffs(1998)]{gal98}
Gallant, Y.\ A., \& Tuffs, R.\ J.\  1998, Mem.\ Soc.\ Astron.\ Ital., 69, 963 


\bibitem[Green \& Scheuer(1992)]{gre92}
Green, D., \& Scheuer, P.\ 1992, MNRAS, 258, 943

\bibitem[Green(2000)]{gre2000}
Green D.\ A.\ 2000, A Catalogue of Galactic Supernova Remnants (2000
August version; Cambridge: MRAO).

\bibitem[Helfand et al.(1995)Helfand, Becker \& White]{hel95} Helfand, D.\ 
J., Becker, R.\ H.\ \& White, R.\ L.\ 1995, \apj, 453, 741 

\bibitem[Hester et al.(1996)]{hes96} Hester, J.\ J.\ et al.\ 
1996, \apj, 456, 225 

\bibitem[Kardashev(1962)]{kar62} Kardashev, N.~S.\ 1962, 
SvA, 6, 317 



\bibitem[Manchester et al.(1993)Manchester, Staveley-Smith, \&
  Kesteven]{man93}
  Manchester, R.\ N., Staveley-Smith, L.\ \& Kesteven, M.\ J.\ 1993, \apj,
  411, 756 

\bibitem[Pacini \& Salvati(1973)]{pac73}
Pacini, F., \& Salvati, M.\ 1973, ApJ, 186, 249

%distance to the center of the Galaxy (R_0) is 8.0 +/- 0.5 kpc
\bibitem[Reid(1993)]{rei93} Reid, M.\ J.\ 1993, \araa,
  31, 345 

\bibitem[Reynolds \& Chanan(1984)]{rey84}
Reynolds, S., \& Chanan, G.\ 1984, ApJ, 281, 673


\bibitem[Safi-Harb et al.(2001)]{saf01} Safi-Harb, S., 
Harrus, I.~M., Petre, R., Pavlov, G.~P., Koptsevich, A.~B., \& Sanwal, D.\ 
2001, ApJ, in press 
\notetoeditor{accepted by ApJ June 22, 2001; to appear in Oct 20, 2001 
issue of ApJ}


% 91 GHz single dish image, 91 and 142 GHz f.d.
\bibitem[Salter et al.(1989a)]{sal89a}
Salter, C.\ J., Emerson, D.\ T., Steppe, H., \& Thum, C.\ 1989a, A\&A,
225, 167.

% Radio spectra of G21 and other Crab-type remnants, 84GHz f.d.
\bibitem[Salter et al.(1989b)]{sal89b}
Salter, C., Reynolds, S., Hogg, D., Payne, J., \& Rhodes, P.\ 1989b, ApJ, 338, 171

\bibitem[Sault et al.(1995)Sault, Teuben, \& Wright]{sau95}
Sault, R.\ J., Teuben, P.\ J., \& Wright, M.\ C.\ H.\ 1995, in ASP Conf.\
Ser. 77, Astronomical Data Analysis Software and Systems IV, ed.
R.\ A.\ Shaw, H.\ E.\ Payne, \& J.\ J.\ E.\ Hayes (San Francisco: ASP), 433

%     AN APPROACH TO INTERFEROMETRIC MOSAICING.
\bibitem[Sault et al.(1996)Sault, Staveley-Smith, \& Brouw]{sau96}Sault, R.\ J., Staveley-Smith, L., \& Brouw, W.\ N. 1996, A\&AS, 120, 375

\bibitem[Slane et al.(2000)]{sla00}
Slane, P., Chen, Y., Schulz, N.\ S., Seward, F.\ D., Hughes, J.\ P., \&
Gaensler, B.\ M.\ 2000, ApJ, 533, L29

\bibitem[Strom \& Greidanus(1992)]{str92}
Strom, R \& Greidanus, H. \ 1992, Nature, 358, 654

\bibitem[Warwick et al.(2001)]{war01} Warwick, R.\ S.\ et 
al.\ 2001, A\&A, 365, L248

%"The Berkeley-Illinois-Maryland-Association Millimeter Array"
\bibitem[Welch et al.(1996)]{wel96}
Welch, W.\ J.\ et al.\ 1996, \pasp, 108, 93

\bibitem[Woltjer et al.(1997)]{wol97} 
Woltjer, L., Salvati, M., Pacini, F., \& Bandiera, R.\ 1997, \aap, 325, 295 


\end{thebibliography}
\end{document}